\documentclass[sigconf]{acmart}

\AtBeginDocument{%
  }

\copyrightyear{2026}
\acmYear{2026}
\setcopyright{cc}
\setcctype{by}
\acmConference[L@S '26] {Proceedings of the Thirteenth ACM Conference on Learning @ Scale}{June 29--July 3, 2026}{Seoul, Republic of Korea.}
\acmBooktitle{Proceedings of the Thirteenth ACM Conference on Learning @ Scale (L@S '26), June 29--July 3, 2026, Seoul, Republic of Korea}
\acmISBN{979-8-4007-2293-6/2026/06}
\acmDOI{10.1145/3774398.3811622}

\usepackage{makecell}
\usepackage{subcaption}
\usepackage{booktabs}
\usepackage{array}
\usepackage{multirow}
\begin{document}

\title{A Large-Scale Observational Study on Obtaining \\Lightweight, Randomized Weekly Student Feedback}
\subtitle{Associations with End-of-Term Course Evaluations}
\renewcommand{\shorttitle}{A Large-Scale Observational Study on Obtaining Lightweight, Randomized Weekly Student Feedback}

\author{Yunsung Kim}
\email{yunsung@stanford.edu}
\orcid{1234-5678-9012}
\affiliation{%
  \institution{Stanford University}
  \city{}
  \state{California}
  \country{USA}
}

\author{Hansol Lee}
\email{hansol@stanford.edu}
\orcid{1234-5678-9012}
\affiliation{%
  \institution{Stanford University}
  \city{}
  \state{California}
  \country{USA}
}

\author{Candace Thille}
\email{cthille@stanford.edu}
\orcid{1234-5678-9012}
\affiliation{%
  \institution{Stanford University}
  \city{}
  \state{California}
  \country{USA}
}

\author{Chris Piech}
\email{cpiech@cs.stanford.edu}
\orcid{1234-5678-9012}
\affiliation{%
  \institution{Stanford University}
  \city{}
  \state{California}
  \country{USA}
}

\newcommand{\note}[1]{\textcolor{blue}{[#1]}}
\newcommand{\todo}{\textcolor{red}{\textbf{TODO}}}
\newcommand{\set}[1]{\left\{#1\right\}}
\newcommand{\paren}[1]{\left(#1\right)}

\begin{abstract}
Student feedback on their course experience provides valuable input for course instructors to reflect on their instruction and make adjustments to their teaching. Conventional methods of obtaining student feedback, however, face a fundamental tradeoff between feedback frequency and quality: as feedback requests become more frequent, participation often declines and responses become less thoughtful over time. To obtain both timely and thoughtful feedback from students, Kim and Piech (\cite{kim2023high}, L@S'23) recently proposed the following simple and lightweight course feedback mechanism: survey each student a small number of times per term, in randomly selected weeks. Named \emph{High-Resolution Course Feedback (HRCF)}, this simple method has been shown to elicit specific and timely feedback that instructors found helpful in understanding their students and making instructional adjustments, without imposing excessive feedback burden on the students.

An important question, however, remains unanswered: is use of this simple method associated with measurable improvements in \emph{students' actual course experiences?} To answer this question, we present a large-scale observational study of HRCF use across 103 course offerings (a total enrollment of 24,216 students) over four years from Fall 2021 through Fall 2025, spanning 42 unique computer science courses at an R1 institution. We analyzed the end-of-term student evaluation data for these courses from Fall 2018 to Fall 2025. Through a regression analysis on the average student ratings on four different student evaluation items, we find that first-time use of HRCF is not associated with a measurable change in average student ratings. However, among small/medium-enrollment course offerings (<250) that continued using HRCF, each additional term of use is associated with average rating increases of 0.045 to 0.048 points for learning-related student evaluation items. We observe no statistically significant associations for large-enrollment course offerings ($\geq250$) or for average ratings of instructional quality and organizational quality of courses. These results provide evidence that repeated HRCF use does have a positive association with increased student evaluations of their learning experience, but that the current design of HRCF needs to be enhanced to support measurable improvements in instruction quality and course organization.
\end{abstract}

\ccsdesc[500]{Applied computing~Computer-assisted instruction}

\keywords{
Course Feedback;
Course Evaluation;
Instructional Improvement;
Weekly Course Feedback;
Student Learning Experience;
Student Course Experience;
Student Evaluations of Teaching;
SETs;
Student Ratings;
End-of-Term Course Evaluations}


\maketitle

\section{Introduction}

The journey of becoming an excellent teacher is a continuous process of reflection and improvement. One of the most widely used inputs for instructional improvement is students' own accounts of their course experiences and their evaluations of an instructor’s teaching, which many instructors use to reflect on their teaching and identify target areas for improvement~\cite{marsh1993use,murray1997does,irons2021enhancing,floden2017impact}. Reflecting this role of student evaluations and feedback, many universities in North America routinely collect student evaluations of teaching to monitor instructional quality and to support instructional improvement~\cite{centra2003will,marsh1997making}.

Student evaluations are most commonly obtained at the end of the academic term. This method, while useful for assessing the quality of instruction, is not well-suited to providing \textit{specific and timely} feedback that supports instructional improvement \textit{during} the course, nor does it benefit the learning experience of the students providing it~\cite{winchester2012if}. To fill in this gap, many instructors additionally collect midterm student evaluations halfway through the course \cite{gravestock2008student,overall1979midterm} or opt to request feedback more frequently on a weekly basis \cite{winchester2012if}. With increased frequency of feedback, a  fundamental tension arises between timeliness and quality. Frequent feedback requests enable a rapid feedback loop but also create excess burden on students such that providing feedback ``soon becomes routine and too onerous a task'' \cite{winchester2012if}, leading to decreased participation and less thoughtful responses over time. This phenomenon is known as ``survey fatigue'' \cite{groves2011survey,porter2004multiple}, and it compromises the validity of the received feedback and results in a systematic nonresponse bias \cite{adams2012nonresponse}.

\begin{figure*}[t]
    \centering
    \includegraphics[width=\linewidth]{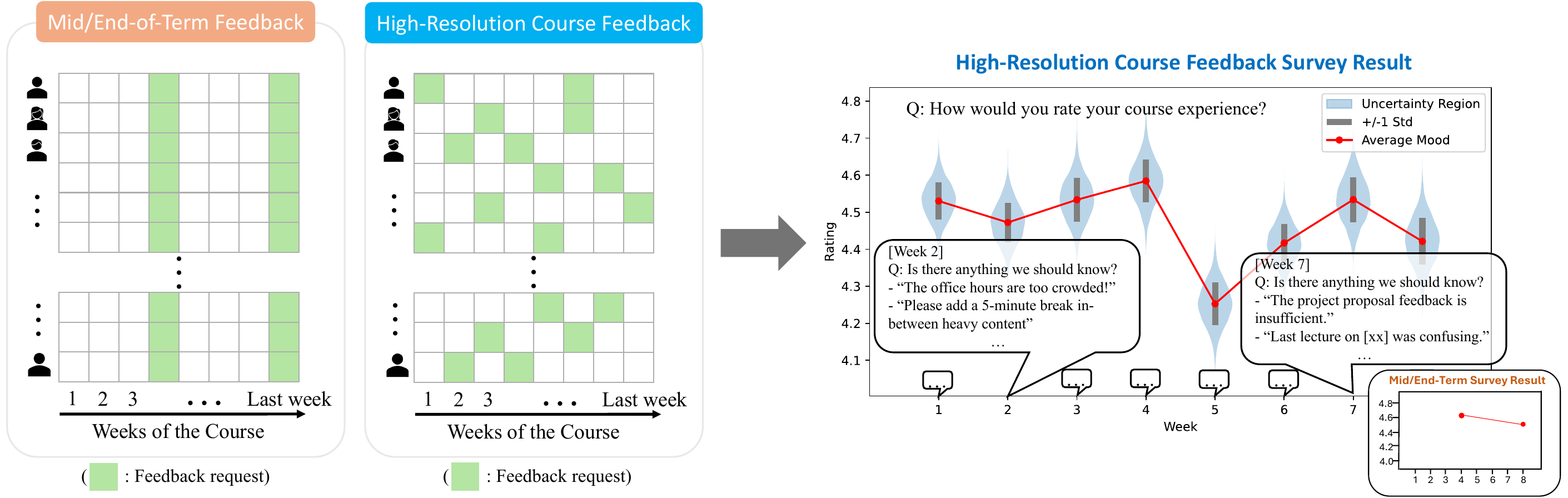}
    \caption{Overview of High-Resolution Course Feedback (HRCF) \cite{kim2023high}. HRCF obtains high-quality and frequent student feedback by randomizing when each student receives feedback, unlike conventional feedback mechanisms that request feedback from all students at a few specific points in the course.}
    \Description{Comparing midterm/end-of-term feedback and High-Resolution Course Feedback. Midterm/End-of-term feedback requests are made at two specific points in the course. HRCF spreads feedback requests across different weeks of the course.}
    \label{fig:hrcf}
\end{figure*}

A simple, lightweight approach to resolving the tradeoff between timely feedback and survey fatigue has recently been proposed. \textbf{``High-Resolution Course Feedback'' (HRCF)}, first proposed at Learning@Scale 2023 \cite{kim2023high}, requests feedback from a small \textit{random} subset of students each week, so that all students are surveyed the same, fixed number of times per term (typically twice) but on random weeks (Figure~\ref{fig:hrcf}). By randomly distributing feedback requests across time, HRCF aims to obtain specific, time-sensitive feedback at a weekly cadence without increasing the burden on students relative to 
conventional feedback mechanisms that survey everyone at specific points in the course. Prior work \cite{kim2023high} found that HRCF generates timely and constructive comments that instructors found useful in understanding the weekly sentiment of students and making adjustments to their instruction.

However, an important aspect of this approach still remains unexplored: whether HRCF use is associated with measurable improvements in \textbf{\textit{students' actual course experiences}}. Whether the effects of the instructional changes made by the instructors meaningfully propagate to students' perceived learning experiences is an important indicator for the instructional value of HRCF and its broader adoption in courses. We conduct a large-scale observational study of HRCF use across four years from Fall 2021 to Fall 2025. During this period a total of 42 computer science courses at an R1 institution used HRCF for a total of 103 course offerings, with 24,216 total enrollments and 15,965 unique students. Using student course evaluation data for the same courses for all offerings between Fall 2018 and Fall 2025, we measured the association between HRCF use and average student ratings for 4 evaluation items (Table~\ref{tab:eot_questions}) regarding (1) the perceived amount of learning goals achieved (GOAL), (2) the perceived learning gains (LEARN), (3) the overall instructional quality (QUAL), and (4) the organizational quality of the course (ORG).

We use regression analyses
to estimate the associations between HRCF use and average student evaluation ratings for each of the four course evaluation items. Specifically, we address the following research questions:
\begin{description}
  \item[RQ1:] Is the use of HRCF associated with students' reported course experiences?
  \item[RQ2:] Is continued use of HRCF across course offerings associated with students’ reported course experiences, and does this marginal association vary by enrollment size?
\end{description}
We find that the initial adoption of HRCF is not associated with a measurable change in average student ratings. However, among small/medium-enrollment ($<250$) course offerings from courses that continue using HRCF, each additional term of HRCF use is associated with average rating increases of $0.045$ (GOAL) and $0.048$ (LEARN) points for learning-related items. We observe no significant associations for large-enrollment course offerings ($\geq250$) or for the QUAL and ORG items for any enrollment size. Taken together, these results provide evidence that repeated HRCF use does have a positive association with improved student evaluations of their learning experience, but that HRCF needs to be enhanced to support broader measurable improvements in instruction quality and course organization.

\section{Related Work}

Educational institutions evaluate teaching effectiveness using a wide range of approaches including ``peer visits, external and internal reviews of teaching portfolios, and qualitative feedback from students,'' but these methods are often more demanding and time-consuming than student evaluations of teaching \cite{read2001relationship}. Consequently, student evaluations of teaching (SETs) have become the most common scalable instrument for monitoring perceived teaching and course quality in universities~\cite{read2001relationship,winchester2012if}. Prior research emphasizes that SETs are used by various educational stakeholders for different purposes~\cite{marsh1993use}: as input to administrative judgments such as tenure~\cite{penny2003changing}; as a channel for students to share past course experiences and obtain information for making enrollment decisions~\cite{marsh2007students,marsh1997making,marsh1993use}; and as signals for instructors and course designers to diagnose issues, iteratively improve courses, and make instructional improvements~\cite{marsh1993use,murray1997does,ebmeier2003supervision,floden2017impact}. High-Resolution Course Feedback is an example of the last ``formative'' use of student feedback.

Decades of prior research on student evaluations and feedback on teaching has explored what these instruments are capable of measuring and how they should be interpreted. Researchers have engaged in active debates about their validity as measures of teaching effectiveness~\cite{abrami1990validity,murray1997does,spooren2013validity,kulik2001student}, their relationship to student learning and academic outcomes~\cite{cohen1981student,murray1997does}, and the extent to which these scores may reflect factors and biases unrelated to teaching effectiveness, such as course structure and size, workload, instructional methods, grading leniency, instructor gender, and student demographic and background characteristics~\cite{feldman1992college,cohen1981student,greenwald1997grading,richardson2005instruments,mckone1999analysis,braskamp1994assessing,centra1993reflective,centra1976relationship,remedios2008liked}. Overall, studies conclude that student evaluations can provide informative signals of instructional quality and actionable directions for improvement~\cite{chen2003student,penny2004effectiveness,murray1997does,marsh1997making} and capture diverse aspects of effective instruction~\cite{spooren2013validity}. Consistent with this, instructors often report that they find student evaluations to be a useful input for refining their teaching practice~\cite{schmelkin1997faculty,arthur2009performativity}.

\begin{table*}[t]
    \centering
    \caption{Default High-Resolution Course Feedback Questions \cite{kim2023high}}
    \label{tab:hrcf_questions}
    \begin{tabular}{lll}
        \toprule
        \textbf{Question Text} & \textbf{Intended Outcome} & \textbf{Response type} \\
        \midrule
        How would you rate your course experience this week? & Overall weekly sentiment & 5-point Likert \\
        What did you like about the course so far? & Positive feedback and encouragement & Free text \\
        Is anything from class still confusing to you? & Self-reflections on the learning process & Free text \\
        Is there anything the teaching team should know? & Fostering open communication & Free text \\
        \bottomrule
    \end{tabular}
\end{table*}

Yet, the quality and utility of student evaluations and feedback depend crucially on the students' willingness to provide careful and authentic feedback. Prior work suggests that this motivation is closely tied to whether students believe that their evaluation and feedback will be valued by their instructors ~\cite{brown2008student,spooren2013validity,lewis2001using,svinicki2001encouraging}. For instance, \cite{winchester2012if} reports that students were more positive towards providing evaluations ``if the lecturer discussed them and/or made changes to their future lectures.'' Similarly, \cite{chen1998assessing} argues that students are motivated to provide evaluations when they see their actions result in a positive outcome. In line with this, \cite{chen2003student} reports that students consider the most desirable outcome of student evaluations of teaching to be ``improvement of teaching,'' with changes to course content and format as a secondary outcome. On the other hand, students often lack the confidence that their feedback will be taken seriously by the instructors, which can lead to apathy and less careful feedback~\cite{spencer2002student}.

In this light, the nature of end-of-term evaluations is largely summative and retrospective, and studies have noted their limitations in benefiting the learning experience of students who are providing the evaluations~\cite{narasimhan2001improving,winchester2012if}. Midterm feedback~\cite{keutzer1993midterm,cohen1981student,abbott1990satisfaction,overall1979midterm} and weekly feedback~\cite{winchester2012if} have been studied as alternatives to end-of-term evaluations for formative purposes. Studies provide evidence that formative feedback can help make improvements to diverse aspects of instruction, such as teaching practice~\cite{chen2003student} as well as course content and structure~\cite{driscoll1979effects,simpson1995uses}. The use of midterm feedback has also empirically been associated with higher end-of-term evaluation ratings~\cite{cohen1981student,overall1979midterm}, improved learning gains and enhanced affective outcomes~\cite{overall1979midterm}, and greater student satisfaction with the feedback process~\cite{veeck2016use,abbott1990satisfaction}. Beyond these measurable outcomes, \cite{diamond2004usefulness} demonstrates that feedback obtained during class can facilitate mutual communication between students and instructors around instructional decision making.

To maximize the benefits of formative feedback on improving teaching quality and student learning experiences, prior work emphasizes the value of eliciting feedback that is both timely and specific \cite{lewis2001using,cross1988classroom,mclaughlin1988teacher}. Weekly feedback was studied as an alternative that can provide continuous formative feedback to the instructors throughout the term \cite{winchester2012if}. When feedback requests were made weekly, however, \cite{winchester2012if} reports that students' willingness to complete the feedback survey varied and noticeably declined over the academic term. When interviewed, students also ``almost unanimously'' did not want to fill out surveys every week noting that students would ``get fed up quickly,'' and several suggested collecting feedback twice a term as a more motivating cadence. High-Resolution Course Feedback \cite{kim2023high} was designed to resolve this tradeoff between feedback timeliness and survey fatigue through randomized sampling.

\section{High-Resolution Course Feedback}

This section provides a brief overview of the High-Resolution Course Feedback mechanism (HRCF) \cite{kim2023high} and preliminary findings about its utility.  HRCF is a formative course feedback mechanism designed to provide course instructors with specific and timely feedback and opportunities to adjust their instruction throughout the course. Instead of collecting feedback from all students at a few fixed points in the course (e.g., midterm and end of term), HRCF makes a fixed number of feedback requests to every student but in randomly selected weeks of the course. As a result, instructors receive feedback from a random subset of students each week. In the baseline setup, each student is invited to respond in $S=2$ randomly chosen weeks, spaced at least two weeks apart.

Operationally, HRCF follows a 3-phase weekly cycle. On Monday of each week, the system emails feedback invitations to the subset of students scheduled for that week with a link to an anonymous survey page. The survey includes the four default questions from Table~\ref{tab:hrcf_questions}, along with optional custom questions of the week set by the instructors. The four default questions are designed to (1) gauge weekly course sentiment, (2) surface encouragement and appreciation for the teaching team, (3) prompt student reflection on their learning process, and (4) support open, ongoing communication between students and instructors. On Thursday, reminder emails are sent to students who have not completed the survey. The week's feedback results are collected and sent to the instructors on Sunday in a weekly digest email. The digest email reports (i) the weekly response rate, (ii) compiled feedback responses with visual summaries for rating items, and (iii) a temporal graph of ``class mood'' ratings calculated as the average ratings of the course experience from the default question (on the scale of ``Excellent,'' ``Good,'' ``Average,'' ``Below Average,'' and ``Poor'').

Prior work has explored the utility of HRCF from the perspective of course instructors. \cite{kim2023high} found that HRCF elicits feedback that is both timely and instructionally useful. In their analysis of weekly HRCF responses from a large, representative course, they found that roughly 25\% to 50\% of comments surfaced ongoing, course-level issues that could be addressed during the term. Instructors reported that HRCF was useful for identifying student pain points and making concrete instructional adjustments to lecture pacing and structure, course content, office-hour logistics, and assignments. They also reported that weekly course experience ratings (“class mood”) were useful for probing the dynamics of course sentiment throughout the course.

\section{Study Method}

\subsection{HRCF Usage Data}

\begin{table}[t]
\centering
\caption{HRCF and End-of-Term Evaluation Dataset Detail}
\label{tab:hrcf_data_stats}
\begin{tabular}{lc}
\toprule
\multicolumn{1}{c}{\textbf{Metric}} & \textbf{Value} \\
\midrule
\textbf{Course Metrics} & \\
\quad Number of Unique Courses & 42 \\
\quad Total Number of Course Offerings & 310 \\
\quad\quad Course Offerings with HRCF & 103 \\
\quad\quad Course Offerings without HRCF & 207 \\

\addlinespace[6pt]
\textbf{HRCF Survey Metrics} & \\
\quad Total Class Size & 24,216 \\
\quad Number of Unique Students & 15,965 \\
\quad Total Feedback Requests & 56,314 \\
\quad Total Responses Received & 22,431 \\
\quad Overall Response Rate & 39.8\% \\

\addlinespace[6pt]
\textbf{End-of-Term Evaluation Metrics} & \\
\quad Average Response Rate & 68.3\% \\
\quad Average Ratings for Each Question & \\
\quad\quad Learning goals (GOAL) & 3.98 \\
\quad\quad How much learned (LEARN) & 4.21 \\
\quad\quad Overall quality (QUAL) & 4.23 \\
\quad\quad How organized (ORG) & 4.14 \\

\bottomrule
\end{tabular}
\end{table}

Our analysis is based on observations of HRCF use primarily within computer science courses at an R1 institution operating on a quarter system. These observations span a total of 13 academic terms from Fall 2021 through Fall 2025. 
Among the courses that used HRCF, we filtered out course offerings with enrollment size smaller than 20, which roughly corresponds to fewer than 4 feedback requests per week on average. This resulted in 103 course offerings across 42 courses that used HRCF, with a total enrollment of 24,216 and 15,965 unique students. The overall HRCF response rate was 39.8\%. Table~\ref{tab:hrcf_data_stats} includes a summary of the HRCF usage data. Among the 42 courses, 21 courses used HRCF in more than one offering. The distribution of course sizes and number of prior HRCF uses per course offering can also be found in Tables~\ref{tab:enroll_dist} and \ref{tab:prior_dist}. 

\begin{table}[t]
\centering
\caption{Distribution of Course Offerings by Enrollment}
\label{tab:enroll_dist}
\begin{tabular}{cccc}
\toprule
\textbf{Enrollment} & \textbf{HRCF} & \textbf{Non-HRCF} & \textbf{Total} \\
\midrule
20--49 & 4 & 12 & 16 \\
50--99 & 11 & 35 & 46 \\
100--199 & 30 & 71 & 101 \\
200--299 & 29 & 51 & 80 \\
300--399 & 16 & 15 & 31 \\
400--499 & 9 & 16 & 25 \\
500+ & 4 & 7 & 11 \\
\midrule
\textbf{Total} & 103 & 207 & 310 \\
\bottomrule
\end{tabular}
\end{table}
\begin{table}[t]
\centering
\caption{Distribution of HRCF Course Offerings by Prior HRCF Use and Enrollment Size ($<250$ vs.\ $\ge250$)}
\label{tab:prior_dist}
\begin{tabular}{cccc}
\toprule
\textbf{Prior HRCF Use} & \textbf{$<$ 250} & \textbf{$\geq$ 250} & \textbf{Total} \\
\midrule
0 (First Time Use) & 35 & 7 & 42 \\
1 & 10 & 11 & 21 \\
2 & 9 & 6 & 15 \\
3 & 4 & 7 & 11 \\
4 & 0 & 5 & 5 \\
5 & 1 & 2 & 3 \\
6 & 0 & 3 & 3 \\
7 & 0 & 1 & 1 \\
8 & 0 & 1 & 1 \\
9 & 0 & 1 & 1 \\
\midrule
\textbf{Total} & 59 & 44 & 103 \\
\bottomrule
\end{tabular}
\end{table}

\subsection{End-of-Term Student Evaluations}

\begin{table*}[t]
    \centering
    \caption{End-of-Term Course Evaluation Questions}
    \label{tab:eot_questions}
    \begin{tabular}{lll}
        \toprule
        \textbf{Label} & \textbf{Question Text} & \textbf{Response type} \\
        \midrule
        GOAL & How well did you achieve the learning goals of this course? & 5-point Likert \\
        LEARN & How much did you learn from this course? & 5-point Likert \\
        QUAL & Overall, how would you describe the quality of the instruction in this course? & 5-point Likert \\
        ORG & How organized was the course? & 5-point Likert \\
        HOURS & How many hours per week on average did you spend on this course (including class meetings)? & Numeric \\
        INPERSON & About what percent of the class meetings (including discussions) did you attend in person? & Percentage \\
        ONLINE & About what percent of the class meetings did you attend online? & Percentage \\
        & What would you like to say about this course to a student who is considering taking it in the future? & Free text \\
        \bottomrule
    \end{tabular}
\end{table*}

For every course that used HRCF at least once, we collected end-of-term student evaluation reports for all course offerings from Fall 2018 through Fall 2025. Evaluation reports between Fall 2018 and Fall 2021 were included to serve as within-course control observations prior to HRCF adoption. This resulted in the inclusion of student evaluation reports for 207 non-HRCF course offerings in addition to the 103 HRCF course offerings. The end-of-term evaluation reports contained student responses to eight questions listed in Table~\ref{tab:eot_questions}. We used the first four items---“Learning goals” (GOAL), “How much learned” (LEARN), “Overall quality of instruction” (QUAL), and “How organized” (ORG)---as our target measures. Ratings used a 5-point Likert scale (1–5). Mean item ratings ranged from 3.98 to 4.23 (Table~\ref{tab:hrcf_data_stats}). Of the remaining items, the three questions on hours spent per week (HOURS) and percent of attendance in-person/online (INPERSON/ONLINE) were used as additional control variables for each course offering to be included in our analysis.

\subsection{Analysis Model}

We use an ordinary least squares (OLS) model to analyze the association between HRCF usage and each of the four target student evaluation ratings. Recall that the purpose of our study is to analyze whether student evaluations are associated with the use of HRCF (\textbf{RQ1}) and whether any association varies with repeated HRCF use and with enrollment size (\textbf{RQ2}). We operationalize these constructs with the following explanatory variables: (1) indicator variable $U_{ct}\in\set{0,1}$ denoting whether course $c$ used HRCF during its offering in term $t$, and (2) variable $K_{ct}\in\set{0,1,2,...}$ denoting the number of prior HRCF uses in the course. We allow the HRCF-use association to change with repeated use and to differ for large-enrollment course offerings.

Our complete regression model is the following:
\begin{equation*}
\begin{aligned}
    &\text{AverageRating}_{ct} = \beta_0 + \delta_c + \delta_y \\
    &\;+ \beta_1\,U_{ct} 
    + \beta_2\left(U_{ct}\cdot K_{ct}\right)
    + \beta_3\left(U_{ct}\cdot K_{ct}\cdot\textbf{1}\{\text{Enrollment}_{ct}\ge 250\}\right) \\
    &\;+ \beta_4(\text{Enrollment})_{ct}
    + \beta_5(\text{Avg Hours Per Week})_{ct} \\
    &\;+ \beta_6(\text{Avg \% of Class Meetings Attended In Person})_{ct} \\
    &\;+ \beta_7(\text{Avg \% of Class Meetings Attended Online})_{ct}.
\end{aligned}
\end{equation*}
In this model, $\beta_1$ captures the association between first-time HRCF use (i.e., when prior use $K_{ct}=0$) and average ratings. We allow the usage effect to vary with repeated use by including the interaction term $\beta_2\paren{U_{ct}\cdot K_{ct}}$. Here, $\beta_2$ represents how the average ratings change per additional prior use. To understand how the association differs in large-enrollment settings, we further include $\beta_3\left(U_{ct}\cdot K_{ct}\cdot\textbf{1}\{\text{Enrollment}_{ct}\ge 250\}\right)$. Coefficient $\beta_3$ measures the additional change in the repeat-use association for large-enrollment course offerings ($\geq$250) relative to smaller course offerings (<250). The enrollment threshold of 250 was chosen to be close to the median enrollment size (225) across HRCF course offerings while maintaining a balanced distribution of repeat-use observations across enrollment bins.\footnote{Using the median as the threshold produces slightly stronger results. We report the more conservative results with 250 as the threshold.}

We control for course-specific ($\delta_c$) and academic year-specific ($\delta_y$) fixed effects. Even within the same course, several factors specific to each offering are known to affect student ratings, including enrollment size, subject area, method of instruction, workload, and grades \cite{mckone1999analysis,braskamp1994assessing,centra1993reflective,centra1976relationship,remedios2008liked}. To the greatest extent possible, we control for these variables using data available to us, which include enrollment size, average hours per week spent (HOURS item in Table~\ref{tab:eot_questions}), and percent of class meetings attended in person and online (INPERSON and ONLINE items in Table~\ref{tab:eot_questions}).

\section{Results and Analysis}

\begin{table*}[t]
\centering
\caption{OLS estimates of associations between HRCF use and course evaluation ratings. Standard errors in parentheses. All models include year and course fixed effects. $^{***}p<0.001$, $^{**}p<0.01$, $^{*}p<0.05$.}
\label{tab:ols_results}
\begin{tabular}{lcccc}
\toprule
 & (1) & (2) & (3) & (4) \\
& \makecell{Learning goals\\(GOAL)}
& \makecell{How much learned\\(LEARN)}
& \makecell{Overall quality\\(QUAL)}
& \makecell{How organized\\(ORG)} \\
\midrule
\addlinespace[6pt]
\textbf{HRCF Effects} & & & & \\
\addlinespace[6pt]
\quad HRCF (first use) & $-0.038$ & $-0.038$ & $-0.043$ & $-0.024$ \\
 & $(0.029)$ & $(0.027)$ & $(0.042)$ & $(0.036)$ \\[2pt]
\quad HRCF Repeat (Enrollment $<$ 250) & $\mathbf{0.044^{*}}$ & $\mathbf{0.048^{*}}$ & $0.046$ & $0.042$ \\
 & $(0.021)$ & $(0.020)$ & $(0.031)$ & $(0.027)$ \\[2pt]
\quad HRCF Repeat (Enrollment $\geq$ 250) & $0.002$ & $0.008$ & $0.003$ & $0.002$ \\
 & $(0.011)$ & $(0.010)$ & $(0.016)$ & $(0.014)$ \\[2pt]

\addlinespace[6pt]
\textbf{Control Variables} & & & & \\
\addlinespace[6pt]
\quad Enrollment (per 100) & $0.034^{*}$ & $0.015$ & $0.085^{***}$ & $0.058^{**}$ \\
 & $(0.017)$ & $(0.016)$ & $(0.025)$ & $(0.022)$ \\[2pt]
\quad Hours per week & $-0.036^{***}$ & $-0.003$ & $-0.037^{***}$ & $-0.029^{**}$ \\
 & $(0.007)$ & $(0.007)$ & $(0.011)$ & $(0.009)$ \\[2pt]
\quad Attend in person (\%) & $0.007^{***}$ & $0.006^{***}$ & $0.012^{***}$ & $0.010^{***}$ \\
 & $(0.002)$ & $(0.002)$ & $(0.002)$ & $(0.002)$ \\[2pt]
\quad Attend online (\%) & $0.006^{***}$ & $0.005^{***}$ & $0.011^{***}$ & $0.010^{***}$ \\
 & $(0.002)$ & $(0.001)$ & $(0.002)$ & $(0.002)$ \\[2pt]
\midrule
\multicolumn{1}{c}{$N$} & 310 & 310 & 310 & 310 \\
\multicolumn{1}{c}{$R^2$} & 0.6183 & 0.7293 & 0.6908 & 0.7045 \\
\multicolumn{1}{c}{Adj.~$R^2$} & 0.5338 & 0.6694 & 0.6224 & 0.6391 \\
\bottomrule
\end{tabular}

\end{table*}

Table~\ref{tab:ols_results} presents the full regression results. We summarize the main technical observations below and provide a more in-depth analysis of these observations in the subsequent sections:

\paragraph{\textbf{First-time HRCF use does not show a statistically significant association with student course evaluations.}} Across all four course evaluation items, no detectable change in average student ratings was observed for first-time HRCF use. 

\paragraph{\textbf{For LEARN and GOAL items, each additional term of prior HRCF use has a statistically significant positive association for small/medium-enrollment course offerings.}} In contrast to first-time use, repeat HRCF use is positively associated with student evaluations in small/medium-enrollment ($< 250$) course offerings with at least one prior HRCF use for ``Learning goals'' ($\beta=0.044^{*}$) and ``How much learned'' ($\beta=0.048^{*}$) items. On the other hand, for large-enrollment courses, the repeat-use associations are not statistically significant. 

\paragraph{\textbf{Repeat HRCF use does not show a statistically significant association for QUAL and ORG items regardless of enrollment size.}} Across these items, repeat HRCF use showed no statistically significant association with average ratings regardless of enrollment size condition.

\subsection{Higher Ratings with Additional Prior HRCF Use in Small/Medium Course Offerings}

Overall, the regression analysis suggests that adopting HRCF for the first time is not associated with measurable changes in students’ end-of-term evaluation ratings. In contrast, among small/medium-enrollment course offerings with prior HRCF use, each additional prior use is associated with meaningfully higher learning-related student ratings. The magnitudes of the estimated regression coefficients for ``Learning goals'' and ``How much did you learn'' items indicate that each additional prior HRCF use is roughly associated with a 0.045- to 0.048-point increase in average course evaluation ratings for small/medium-enrollment (<250) course offerings from courses that continue to use HRCF. In practical terms, this means that, within the same course, an offering after two prior HRCF uses would be predicted to have GOAL/LEARN ratings roughly 0.1 points higher than the course’s first HRCF offering. Considering that average course ratings are already near or above 4 on a 5-point scale (see Table~\ref{tab:hrcf_data_stats}) and that HRCF surveys are low-intensity (two short surveys per student), these associations should be interpreted as meaningful improvements in student perceptions of their learning. 

The positive repeat-use association for GOAL/LEARN items in small/medium-enrollment course offerings fits prior observations that HRCF provides useful formative feedback for making instructional improvements that address difficulties in student learning. The HRCF question ``Is anything from class still confusing to you?'' is specifically targeted at providing instructors with an understanding of which course content students need the most support with. In addition to this question, instructors reported using this feedback for making instructional changes that specifically address student pain points in learning \cite{kim2023high}, such as adding short recaps and clarifications in the beginning of lectures, providing more worked examples during lecture, allocating more time to address commonly confusing course content, providing additional learning materials or practice problems, and restructuring office hours. A finer-grained knowledge of content-specific student difficulties along with instructor-reported adjustments made to improve learning experiences offers a plausible mechanism for the observed association.

\subsection{Learning Curve and First-Time Use}

The null associations of initial adoption, when taken together with the positive repeat-use association, also align with long-running findings within the student evaluations literature that improvements in teaching require deliberate sense-making and structured follow-through in addition to merely collecting feedback \cite{penny2004effectiveness,lewis2001using}. It is a well-known concept in feedback research more generally that individuals vary in the capacities, dispositions, and prior experiences that shape how well they can interpret and use feedback in learning~\cite{carless2018development}. The same insight applies to instructors using feedback to learn about their instructional effectiveness and make improvements, and not all instructors have the capacity to process student feedback and translate it into effective instructional adjustments \cite{cohen1980effectiveness}. This could explain why first-time uses showed no statistically significant association, as instructors could have been facing a learning curve where they were initially unfamiliar with how to respond to student feedback.

This is also particularly relevant in the context of our current study design. Our study is an observational study that leverages naturally occurring variations in HRCF adoption and repeated use across academic terms to measure its association with student evaluations. Because the choice to adopt HRCF and continue using it is at the discretion of the instructor, it is plausible that courses that used HRCF more than once may partially reflect contexts in which instructors were more proficient in processing and interpreting student feedback to make effective instructional adjustments. This could further explain the positive associations with student ratings for small/medium-enrollment course offerings with prior HRCF use. These results suggest that structured support for helping instructors make meaningful interpretations of the feedback they receive and guide them in deriving effective instructional improvement decisions could help improve student learning experiences.

\subsection{Constraints of Acting on Feedback at Scale}

On the other hand, our analysis finds neither first-time nor repeated HRCF use to be associated with any student evaluation item in large-enrollment course offerings. This result is consistent with prior work on the practical constraints of large-enrollment instruction. Large courses face structural challenges of scale that small to medium-enrollment courses do not. These structural challenges include reduced opportunities for student–instructor interaction, limited capacity for individualized student support, greater classroom management and coordination demands, constraints on course structure and delivery, difficulties providing timely and high-quality assessment and feedback at scale, and greater heterogeneity in student preparedness \cite{mulryan2010teaching,cuseo2007empirical,monks2011impact}. These factors not only negatively affect student-reported learning and evaluations of teaching quality in large-enrollment courses \cite{monks2011impact}, but also make instructional adaptation more difficult \cite{mulryan2010teaching}. This rigidity is further compounded by the additional staff time and resources required to design, implement, and coordinate such changes \cite{mulryan2010teaching}. Even when the instructors do make instructional improvements, the constraints of scale may attenuate how much change can propagate into noticeable differences for individual learners. Further research should investigate how the method and practice of receiving and delivering lightweight, weekly feedback could be improved with a focus on making tangible enhancements to the everyday learning experiences of individual learners.

\subsection{Limited Associations with Global Composite Metrics of Course Quality}

Unlike with LEARN and GOAL items, neither first-time nor repeat use had statistically significant associations with the ratings of the overall quality of instruction (QUAL) and organizational quality (ORG) regardless of enrollment size. These results suggest that the effect of the instructional adjustments that HRCF supports may be localized to students' perceptions of learning as opposed to broader perceptions of quality of instruction and course organization. Prior work on student evaluations of teaching emphasizes that composite evaluations such as evaluations of instruction and course quality often reflect various course-related factors other than learning and instruction alone \cite{spooren2013validity}, such as course design, assessment and grading practices, fairness of grading, grading leniency, and workload dynamics  \cite{remedios2008liked,centra2003will}. Many of these structural factors, such as course design and schedule as well as workload, are typically determined in advance and are less amenable to change compared to instructional practices. Consistent with this interpretation, a recent study \cite{levinsson2024course} finds that features of course design are stronger predictors of students’ perceived course quality than instructor-related factors, suggesting that composite metrics like QUAL and ORG may be less responsive to incremental instructional changes during a term that the current design of High-Resolution Course Feedback seems to mainly provide. How High-Resolution Course Feedback can be improved to support meaningful enhancements in broader, structural elements of a course remains an interesting direction for future work.

\section{Limitations and Future Work}

Our findings should be interpreted in light of several limitations of the data and study design. First, our outcomes are based on end-of-term student evaluation ratings. While student evaluations can be meaningful measures of downstream impact of instructional improvements on student learning experiences, this outcome measure misses many other aspects of instructional improvement such as objective learning outcomes or alignment with known effective instructional practices. Other meaningful measures of HRCF's effectiveness include improvements in students' actual learning gains, evaluations from teaching specialists, peer evaluations, and indicators of student engagement, motivation, and confidence. Future work should consider a more comprehensive evaluation of the quality of instructional improvements enabled by HRCF and measure their impact on a diverse set of metrics.

Another limitation of our study is generalizability across domains. Our study relies on data collected from a single discipline (computer science) at a single R1 institution. While analyzing courses within a single discipline allows for more consistent comparisons across course offerings, it also narrows the contexts in which our findings can generalize to. Other disciplines with different pedagogical practices (e.g., studio- or lab-based instruction, writing-intensive instruction), different assessment structures, or different expectations around student-instructor interactions may result in different contexts and constraints of acting on HRCF feedback. Likewise, different institutional policies, teaching support resources, enrollment size distributions, and student cultures can shape both the types of feedback that students provide and the instructors' capacity to process and respond to feedback. More work is needed to collect evidence of the effects of HRCF use to provide a more generalizable understanding of its effectiveness.

Our results suggest that instructors first adopting HRCF may be facing a learning curve, and instructors who are proficient at processing and responding to feedback may benefit more from the tool compared to others who do not. Moreover, the associations observed for large courses may be limited by issues of scale. Future work should explore targeted enhancements to HRCF that support instructors' sense-making of the received feedback and make an impact in large courses. Concretely, HRCF may incorporate automated clustering and prioritization of recurring issues or use AI to generate helpful action plans that would support instructional improvements based on the received feedback. Additionally, since students' motivation to provide thoughtful feedback depends on their belief that their feedback will be valued, HRCF could be enhanced to support an interface where instructors respond directly to students' feedback. Implementing and evaluating these improvements via field experiments would be an important area of future work.


\section{Conclusion}

We studied the associations between using High-Resolution Course Feedback, a simple and lightweight feedback mechanism for obtaining timely and actionable feedback from students, and students' actual learning experiences as measured by course evaluation ratings. High-Resolution Course Feedback obtains weekly feedback from students without creating survey fatigue by requesting feedback from a random sample of students each week, with every student getting the same, small number of feedback requests throughout the term. Through a large-scale observational study of 42 unique computer science courses and 103 course offerings that used HRCF over a 4-year span in an R1 institution, we analyzed the association between HRCF use and student evaluation ratings of their course experience, and whether this association varied by repeated use and enrollment size. We find that, while first-time HRCF use is not associated with a measurable change in student evaluation ratings, each additional use of HRCF has a statistically significant association with a 0.045- to 0.048-point increase in average ratings for learning-related evaluation items for course offerings with small/medium-enrollment (<250). No statistically significant associations were observed for large-enrollment ($\geq250$) course offerings, and for evaluations of overall instructional quality and organizational quality. Overall, the results suggest that HRCF is a useful mechanism associated with improvements in students' perceived learning, but that it needs to be improved to have a broader impact on students' overall course experience. Future work should explore effective methods for supporting instructors' sense-making of the received weekly feedback and consider enhancements of HRCF that would result in measurable improvements in overall quality of instruction and course organization.

\bibliographystyle{ACM-Reference-Format}
\bibliography{sample-base}

\end{document}